\documentclass[twocolumn,aps,prl,floatfix,superscriptaddress]{revtex4-1}
\usepackage{amsmath,amssymb}
\usepackage{graphicx,float}
\usepackage{times,txfonts}
\usepackage{color}
\usepackage{multirow}
\usepackage{bbold}
\usepackage{color}
\usepackage{soul}
\usepackage{hyperref}
\usepackage{times,txfonts}
\usepackage{natbib}
\usepackage{nicefrac}
\usepackage{blindtext}
\usepackage[export]{adjustbox}
\usepackage{mathrsfs}
\usepackage{textcomp}
\usepackage{blkarray}
\usepackage{multirow}
\usepackage{sistyle}
\usepackage{soul}
\definecolor{lightblue}{RGB}{185,210,248}
\sethlcolor{lightblue}

\begin{document}

\title{Performance analysis of table-top single-pulse terahertz detection up to 1.1~MHz}

\author{Nicolas Couture}
\affiliation{Department of Physics, University of Ottawa, 25 Templeton, Ottawa, Ontario, K1N 6N5, Canada}
\affiliation{Max Planck Centre for Extreme and Quantum Photonics, Ottawa, Ontario K1N 6N5, Canada}

\author{Markus Lippl}
\affiliation{Max Planck Institute for the Science of Light, Erlangen 91058, Germany}
\affiliation{Department of Physics, University of Erlangen-Nürnberg, Erlangen 91058, Germany}

\author{Wei Cui}
\affiliation{Department of Physics, University of Ottawa, 25 Templeton, Ottawa, Ontario, K1N 6N5, Canada}
\affiliation{Max Planck Centre for Extreme and Quantum Photonics, Ottawa, Ontario K1N 6N5, Canada}

\author{Angela Gamouras}
\affiliation{National Research Council Canada, Ottawa, Ontario K1A 0R6, Canada}
\affiliation{Department of Physics, University of Ottawa, 25 Templeton, Ottawa, Ontario, K1N 6N5, Canada}

\author{Nicolas Y. Joly}
\affiliation{Department of Physics, University of Erlangen-Nürnberg, Erlangen 91058, Germany}
\affiliation{Max Planck Institute for the Science of Light, Erlangen 91058, Germany}
\affiliation{Interdisciplinary Center for Nanostructured Films, Erlangen 91058, Germany}

\author{Jean-Michel M\'enard}
\affiliation{Department of Physics, University of Ottawa, 25 Templeton, Ottawa, Ontario, K1N 6N5, Canada}
\affiliation{Max Planck Centre for Extreme and Quantum Photonics, Ottawa, Ontario K1N 6N5, Canada}
\affiliation{National Research Council Canada, Ottawa, Ontario K1A 0R6, Canada}
\email{jean-michel.menard@uottawa.ca}

\begin{abstract}
Slow data acquisition in terahertz time-domain spectroscopy (THz-TDS) has hindered the technique’s ability to resolve “fast” dynamics occurring on the microsecond timescale. This timescale, arguably too slow to be accessed via standard optical pump-probe techniques relying on ultrafast sources, hosts a range of phenomena that has been left unexplored due to a lack of proper real-time monitoring techniques. In this work, chirped-pulse spectral encoding, a photonic time-stretch technique, and high-speed electronics are used to demonstrate time-resolved THz detection at a rate up to 1.1~MHz. This configuration relies on a table-top source and a setup able to resolve every THz transient that it can generate. We investigate the performance of this system at different acquisition rates in terms of experimental noise, dynamic range, and signal-to-noise ratio. Our results pave the way towards single-pulse THz-TDS at arbitrarily fast rates to monitor complex dynamics in real-time.
\end{abstract}

\maketitle

Terahertz time-domain spectroscopy (THz-TDS) relies on resolving the oscillating electric field of a THz pulse in order to access its frequency components via Fourier transform. This technique provides full amplitude and phase information of the light passing through a medium, allowing the complex dielectric function of the medium to be extracted without the need for Kramers-Kronig relations, a powerful capability compared to other spectroscopy techniques monitoring only the transmitted optical power. THz-TDS is often performed with a detection technique involving the mechanical scanning of an ultrashort near-infrared (NIR) pulse across the THz waveform as the two interact in a nonlinear crystal. Because this technique intrinsically relies on the acquisition of multiple data points to reconstruct the full THz waveform, it requires a sample under study to exhibit the same characteristics every time it is probed by the THz wave. Thus, standard pump-probe is not a viable technique in evaluating samples whose properties evolve chaotically or experience irreversible changes. Many successful studies have tackled this issue by enabling single-shot THz detection, eliminating the need for a mechanical delay line to retrieve the time-domain THz waveform. These techniques, however, require multiple scans to achieve a high signal-to-noise ratio (SNR). These detection schemes can rely on echelon mirrors~\cite{minami_single-shot_2013}, chirped-pulse spectral encoding~\cite{jiang_electro-optic_1998}, or spectral interferometry~\cite{sharma_terahertz_2012}, and have enabled kHz detection rates~\cite{teo_invited_2015}. In fact, the combination of chirped-pulse spectral encoding and a photonic time-stretch technique~\cite{goda_theory_2009,goda_dispersive_2013}, where the repetition rate of the NIR source used for spectral encoding sets the acquisition rate, has resulted in MHz detection rates~\cite{roussel_observing_2015,szwaj_high_2016,evain_direct_2017,steffen_compact_2020,roussel_phase_2022}. However, these MHz rates experiments were achieved with high-energy THz pulses from large synchrotron facilities to achieve a satisfactory SNR. Nevertheless, this approach has also permitted table-top THz-TDS at a rate of 50~kHz using a single ultrafast source for THz generation and detection, resolving pulse-to-pulse microsecond carrier dynamics in a semiconductor~\cite{couture_single-pulse_2023}. Reaching faster THz-TDS rates with table-top sources would allow complex dynamics at sub-microsecond timescales to be recorded and significantly expedite the data acquisition process in experiments such as THz two-dimensional spectroscopy~\cite{gao_high-speed_2022}. To establish the suitability of a new THz-TDS scheme for such applications, the dynamic range and SNR must be thoroughly investigated. In this work, we use a single-pulse detection technique employing chirped-pulse spectral encoding and a photonic time-stretch technique to resolve THz waveforms up to a rate of 1.1~MHz, relying only on a single ultrafast source. The dynamic range and SNR of each measurement is studied to highlight the feasibility of the scheme for spectroscopy purposes. The promising results we present here lay the foundation for table-top THz-TDS at high acquisition rates as a real-time monitoring tool.

\begin{figure}[!t]
\centering
\includegraphics[width=8 cm]{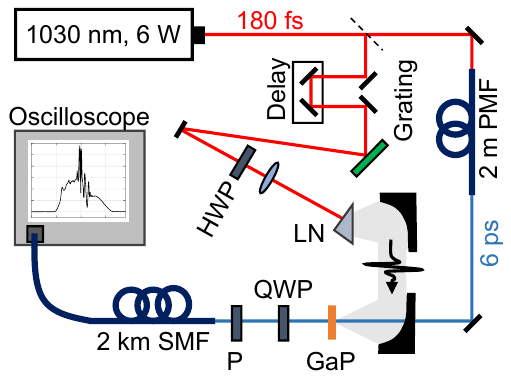}
\caption{\textbf{Experimental setup.} A Yb:KGW amplifier with 6 W average power and tunable repetition rate is employed for THz generation and detection. Most of the optical power is used for THz generation with a tilted-pulse-front technique in a lithium niobate (LN) crystal wedge. The remainder is launched into a 2~m-long polarization-maintaining fiber (PMF) to generate a chirped supercontinuum (SC) with 100~nm bandwidth and 6~ps time duration (FWHM). The THz pulse and the chirped SC are overlapped in a 2~mm-thick gallium phosphide (GaP) crystal to achieve chirped-pulse spectral encoding, imprinting the THz waveform onto the chirped NIR spectrum through the Pockels effect. After polarization filtering with a quarter-wave plate (QWP) and linear polarizer (P), photonic time-stretch is realized by injecting the THz-modulated SC into a 2~km-long single-mode fiber (SMF) and detecting it with a fast photodiode (12~GHz) and oscilloscope (8~GHz).}
\label{fig:setup}
\end{figure}

For these experiments, an amplified ultrafast source centered at a wavelength of 1030~nm delivers 180~fs pulses for THz generation and detection. The laser is operated at its maximum average power of 6~W and its repetition rate is modified via software between 1~kHz to 1.1~MHz, altering only the output peak intensity while chirp and pulse duration remain effectively unaffected. The majority of the output power (90\%) is used for THz generation via optical rectification in a lithium niobate crystal with the tilted-pulse-front technique to ensure an efficient generation process and relatively high THz electric fields~\cite{hebling_generation_2008}. The rest of the beam is set to a constant $\sim$10~nJ pulse energy and launched into a 2~m-long polarization-maintaining fiber (PMF, OZ Optics PMF-980-6/125-0.25-L) to achieve a chirped NIR supercontinuum (SC) spanning over $\sim$100 nm with a time duration of 6~ps (FWHM). A chirped SC with these specifications allows for frequencies up to 1.6~THz to be detected through chirped-pulse spectral encoding, imprinting the THz time-domain waveform onto the chirped NIR spectrum through nonlinear effects~\cite{jiang_electro-optic_1998,couture_single-pulse_2023}. This is achieved by overlapping the resulting THz pulse and chirped SC in a 2~mm-thick 110-oriented gallium phosphide (GaP) crystal. The NIR pulse containing the THz information is transmitted through a quarter-wave plate and linear polarizer to optimize detection sensitivity while maintaining the phase information of the THz pulse~\cite{jiang_electro-optic_1999}. Finally, the encoded NIR pulse is launched into a 2~km-long single-mode fiber (SMF, Corning HI1060 flex) to achieve photonic time-stretch, dispersing the pulse duration from a few picoseconds to tens of nanoseconds, which can then be sampled with a high-speed photodiode (12~GHz bandwidth, Newport 1544-B) and oscilloscope (8~GHz bandwidth, Tektronix MSO 64B). A diagram of the experimental configuration is shown in Fig.~\ref{fig:setup}. With this technique, the THz detection rate can be arbitrarily high and is determined solely by the repetition rate of the ultrafast source providing NIR pulse energies of at least a few $\mu$J.

\begin{figure}[!ht]
\centering
\includegraphics[width=8 cm]{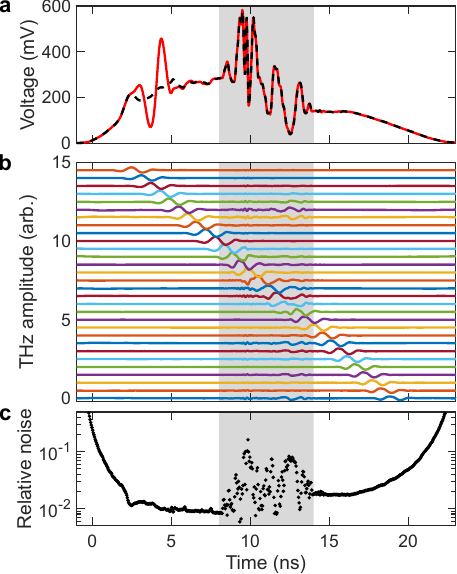}
\caption{\textbf{Time-axis calibration.} a) Unmodulated (dashed black line) and THz-modulated (red line) time-stretched signals measured with the fast photodiode and oscilloscope. To isolate the THz waveform, the square root of the unmodulated signal is subtracted from the square root of the THz-modulated signal. b) The extracted THz waveforms as the relative delay between the THz and chirped SC is shifted in increments of 500 fs, varying the frequencies within the SC that the THz transient is imprinted onto. The waveforms are stacked vertically for clarity. The shaded grey area indicates noisy parts of the spectrum and is quantified by c) the relative noise of the SC, which we define as the standard deviation of the SC divided by the mean SC signal.}
\label{fig:delay dep}
\end{figure}

To retrieve the THz waveform through the photonic time-stretch technique, the signal is recorded on the oscilloscope with and without the THz pulse impinging on the GaP crystal, as shown in Fig.~\ref{fig:delay dep}a (red line and dashed black line, respectively). Subtracting the square root of each measurement yields the THz waveform in the time-stretch domain~\cite{couture_single-pulse_2023}. The recovered waveform is presented in Fig.~\ref{fig:delay dep}b when the delay between the chirped SC and THz pulse is varied by increments of 500 fs. By imprinting the THz on a different portion of the NIR spectrum, the time-axis of the oscilloscope can be calibrated to retrieve the picosecond information of the THz pulse. The linear relationship between the THz peak measured on the oscilloscope and relative delay allows us to extract the time-stretch factor of 1138 and confirms that higher-order dispersion in the 2~km-long SMF is negligible~\cite{goda_theory_2009}. The grey highlighted areas in Fig.~\ref{fig:delay dep} correspond to the portion of the spectrum where the THz waveforms are subject to deformations, and that we have therefore deemed suboptimal for spectral encoding. This spectral region is sensitive to intensity and fiber-coupling fluctuations whereas the rest of the spectrum has inherently higher stability, as displayed by the relative noise of the SC signal shown in Fig.~\ref{fig:delay dep}c. Delaying the pulses such that the THz waveform is imprinted on the most stable parts of the NIR spectrum is a valid method of performing these experiments as the THz information is only contained within $\sim$4 ns and the time-stretched spectrum spans $>$20 ns. The noise recorded in these most stable regions is solely limited by electronic noise from the oscilloscope and pulse-to-pulse laser power fluctuations. To use the entire SC spectrum for spectral encoding, balanced detection techniques adapted to single-pulse detection, such as diversity electro-optic sampling (EOS)~\cite{roussel_phase_2022}, can be used to reduce pulse-to-pulse fluctuations.

\begin{figure}[!t]
\centering
\includegraphics[width=8 cm]{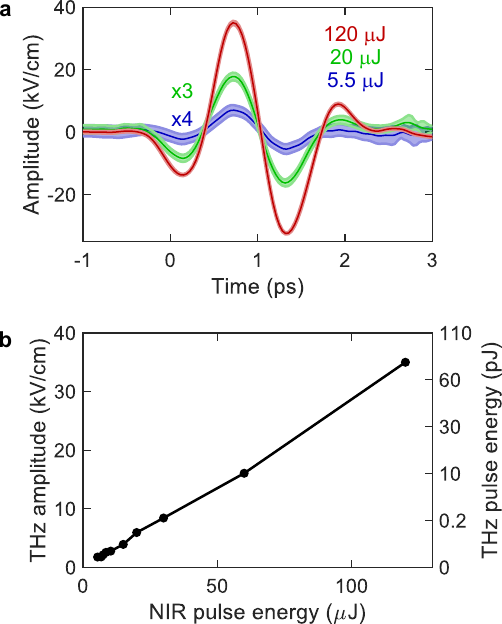}
\caption{\textbf{Relative amplitude characterization.} a) The extracted THz transients generated with NIR pulse energies of 120~$\mu$J (red), 20~$\mu$J (green), and 5.5~$\mu$J (blue); corresponding to detection rates of 50~kHz, 300~kHz, and 1.1~MHz, respectively. The shaded area represents the error of the measurement calculated from the standard deviation over 10k pulses. For clarity, the blue and green lines (and their corresponding standard deviation) are multiplied by factors of 3 and 4, respectively. b) The peak THz transient amplitude as the pulse energy of the ultrafast source is increased from 5.5~$\mu$J to 120~$\mu$J by decreasing the repetition rate (i.e. 1.1~MHz to 50~kHz). The linear relationship between detected THz amplitude and NIR pulse energy indicates that the system can reach higher detection rates at the cost of detection efficiency. The highest THz field, corresponding to a detection rate of 50~kHz, is $\sim$35~kV/cm or $\sim$85~pJ.}
\label{fig:rep rate}
\end{figure}

Figure~\ref{fig:rep rate}a displays the extracted time-domain THz waveform after performing the time-axis calibration for laser output pulse energies of 120, 20, and 5.5~$\mu$J, corresponding to laser repetition rates of 50~kHz, 300~kHz, and 1.1~MHz, respectively. The field strength and pulse energy of the THz pulses, in kV/cm and pJ, respectively, are extracted with EOS and not the single-pulse detection configuration~\cite{blanchard_generation_2007}. For clarity, the data in Fig.~\ref{fig:rep rate}b collected with a 20~$\mu$J and 5.5~$\mu$J NIR pulse energies are multiplied by factors of 3 and 4, respectively. The shaded areas surrounding the colored lines in Fig.~\ref{fig:rep rate}a represent the standard deviation measured over 10k pulses.  Although the signal at 5.5~$\mu$J (1.1~MHz) is weaker, this result marks, to our knowledge, the fastest table-top time-resolved THz detection rate to date.

\begin{figure}[!hb]
\centering
\includegraphics[width=8 cm]{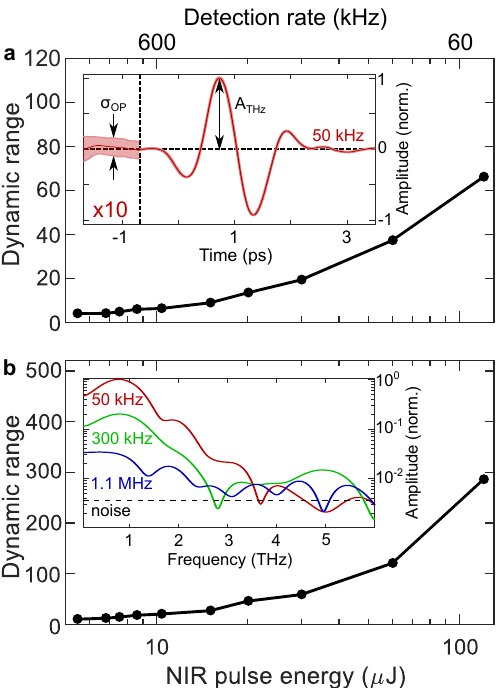}
\caption{\textbf{Dynamic range of single-pulse THz detection.} The dynamic range of the presented scheme plotted as a function of the detection rate (i.e. NIR pulse energy). The dynamic range of a) the recorded time-domain data and b) corresponding THz spectra are calculated with the methods described in Ref.~\cite{naftaly_methodologies_2009}. The arrows in the inset of a) indicate parts of the THz waveform used to calculate the dynamic range. The inset of b) contains THz spectra (plotted in log scale) recorded at repetition rates of 50~kHz (red), 300~kHz (green), 1.1~MHz (blue), and the noise floor of the single-pulse case (dashed black line).}
\label{fig:DR}
\end{figure}

The THz waveform is measured for several pulse energies to form the line plotted in Fig. 3b, where the peak THz amplitude and corresponding pulse energy is shown as a function of the laser output pulse energy. The linear relationship between the NIR pulse energy and the detected THz amplitude validates the linearity of the polarization filtering scheme in Fig.~\ref{fig:setup}~\cite{couture_single-pulse_2023}. For each measurement, the pulse energy injected into the PMF is kept fixed at $\sim$10 nJ, hence, the NIR pulse energy in the GaP detection crystal is also fixed, indicating that the detection efficiency is not limited by the repetition rate but instead the THz field strength inside the crystal. The highest THz field in this work, corresponding to the red line in Fig.~\ref{fig:rep rate}a, is $\sim$35~kV/cm and corresponds to a THz pulse energy of $\sim$85~pJ. In contrast to other schemes, since our system only relies on a single laser and a single photodiode, timing jitter is negligible. Jitter between the THz and the SC would appear in the experimental data as phase noise. In this work, the standard deviation of the phase is on the order 10$^{-2}$ radians across the whole spectrum. This feature makes it highly suitable as a non-invasive probe in industrial assembly lines. The temporal shift between THz and SC induced by a product in the THz path can be used to extract the thickness of the product with great accuracy.

For a more in-depth analysis of the measured THz waveforms at each of the studied repetition rates, we calculate the dynamic range of the recorded time-domain signals (Fig.~\ref{fig:DR}a) and their corresponding spectral amplitude (Fig.~\ref{fig:DR}b). The dynamic range in the time-domain is defined as the mean of the peak THz amplitude ($A_{THz}$) divided by the off-peak standard deviation ($\sigma_{OP}$), whereas in the Fourier domain it is defined as the maximum spectral amplitude of a single-pulse measurement divided by the noise floor (dashed black line inset Fig.~\ref{fig:DR}b)~\cite{naftaly_methodologies_2009}. Notably, the peak dynamic range approaches 300 in amplitude (50~dB in power) in the Fourier domain when operating the system at 50 kHz. The SNR of the single-pulse data can be extracted with a similar approach. In both domains, the SNR is defined as the quotient between the peak THz amplitude and the on-peak standard deviation~\cite{naftaly_methodologies_2009}. The peak SNR achieved with a repetition rate of 50~kHz is $\sim$60 in the time domain and $\sim$150 in the Fourier domain. The dynamic range in each case follows the same trend: As the repetition rate is increased, the dynamic range decreases correspondingly. This trend is a result of the weaker nonlinear interactions in the GaP detection crystal and the constant noise floor. At repetition rates exceeding 600~kHz, the frequency domain SNR of the single pulse measurement approaches unity and is therefore too low to perform any kind of spectroscopy without pulse-to-pulse averaging. Nonetheless, the fact that single THz pulses can be recorded at MHz repetition rates with this scheme is promising for future applications of single-pulse THz-TDS at high repetition rates. For example, an oscillator delivering tens of $\mu$J NIR pulses would allow the presented system to investigate single-pulse dynamics approaching the nanosecond timescale.


In summary, we have used chirped-pulse spectral encoding and a photonic time-stretch technique to demonstrate time-resolved THz detection at a rate up to 1.1~MHz using a single ultrafast source, to our knowledge, the fastest single-pulse table-top detection rate to date. By thoroughly investigating the noise of the presented system, we have deduced the limitation of the system to be the THz field strength at high repetition rates and, hence, low NIR pulse energies. With the simple addition of high-speed electronics and commercially available optical fibers, existing systems with high THz fields (tens of kV/cm) can almost effortlessly implement the presented detection scheme. We believe that this work leads the way towards table-top THz-TDS with high sensitivity to resolve sub-microsecond dynamics in exotic systems that are evolving on a pulse-to-pulse basis.

\vspace{0.5 EM}
\noindent\textbf{Funding} J.-M.M. acknowledges funding from the Natural Sciences and Engineering Research Council of Canada (NSERC) (RGPIN-2023-05365 and RTI-2023-00252) and the Canada Foundation for Innovation (CFI) (Project Number 35269). N.C. acknowledges financial support from the Ontario Graduate Scholarship. M.L. is part of the Max Planck School of Photonics supported by BMBF, Max Planck Society, and Fraunhofer Society. N.Y.J. and M.L. acknowledge the Max Planck Institute for the Science of Light in Erlangen for financial support. This work was also supported by the National Research Council of Canada via the Joint Centre for Extreme Photonics (JCEP).  

\noindent\textbf{Disclosures} The authors declare no conflicts of interest.

\bibliography{Manuscript}

\begin{thebibliography}{17}%
\makeatletter
\providecommand \@ifxundefined [1]{%
 \@ifx{#1\undefined}
}%
\providecommand \@ifnum [1]{%
 \ifnum #1\expandafter \@firstoftwo
 \else \expandafter \@secondoftwo
 \fi
}%
\providecommand \@ifx [1]{%
 \ifx #1\expandafter \@firstoftwo
 \else \expandafter \@secondoftwo
 \fi
}%
\providecommand \natexlab [1]{#1}%
\providecommand \enquote  [1]{``#1''}%
\providecommand \bibnamefont  [1]{#1}%
\providecommand \bibfnamefont [1]{#1}%
\providecommand \citenamefont [1]{#1}%
\providecommand \href@noop [0]{\@secondoftwo}%
\providecommand \href [0]{\begingroup \@sanitize@url \@href}%
\providecommand \@href[1]{\@@startlink{#1}\@@href}%
\providecommand \@@href[1]{\endgroup#1\@@endlink}%
\providecommand \@sanitize@url [0]{\catcode `\\12\catcode `\$12\catcode
  `\&12\catcode `\#12\catcode `\^12\catcode `\_12\catcode `\%12\relax}%
\providecommand \@@startlink[1]{}%
\providecommand \@@endlink[0]{}%
\providecommand \url  [0]{\begingroup\@sanitize@url \@url }%
\providecommand \@url [1]{\endgroup\@href {#1}{\urlprefix }}%
\providecommand \urlprefix  [0]{URL }%
\providecommand \Eprint [0]{\href }%
\providecommand \doibase [0]{http://dx.doi.org/}%
\providecommand \selectlanguage [0]{\@gobble}%
\providecommand \bibinfo  [0]{\@secondoftwo}%
\providecommand \bibfield  [0]{\@secondoftwo}%
\providecommand \translation [1]{[#1]}%
\providecommand \BibitemOpen [0]{}%
\providecommand \bibitemStop [0]{}%
\providecommand \bibitemNoStop [0]{.\EOS\space}%
\providecommand \EOS [0]{\spacefactor3000\relax}%
\providecommand \BibitemShut  [1]{\csname bibitem#1\endcsname}%
\let\auto@bib@innerbib\@empty
\bibitem [{\citenamefont {Minami}\ \emph {et~al.}(2013)\citenamefont {Minami},
  \citenamefont {Hayashi}, \citenamefont {Takeda},\ and\ \citenamefont
  {Katayama}}]{minami_single-shot_2013}%
  \BibitemOpen
  \bibfield  {author} {\bibinfo {author} {\bibfnamefont {Y.}~\bibnamefont
  {Minami}}, \bibinfo {author} {\bibfnamefont {Y.}~\bibnamefont {Hayashi}},
  \bibinfo {author} {\bibfnamefont {J.}~\bibnamefont {Takeda}}, \ and\ \bibinfo
  {author} {\bibfnamefont {I.}~\bibnamefont {Katayama}},\ }\href {\doibase
  10.1063/1.4817011} {\bibfield  {journal} {\bibinfo  {journal} {Appl. Phys.
  Lett.}\ }\textbf {\bibinfo {volume} {103}},\ \bibinfo {pages} {051103}
  (\bibinfo {year} {2013})}\BibitemShut {NoStop}%
\bibitem [{\citenamefont {Jiang}\ and\ \citenamefont
  {Zhang}(1998)}]{jiang_electro-optic_1998}%
  \BibitemOpen
  \bibfield  {author} {\bibinfo {author} {\bibfnamefont {Z.}~\bibnamefont
  {Jiang}}\ and\ \bibinfo {author} {\bibfnamefont {X.-C.}\ \bibnamefont
  {Zhang}},\ }\href {\doibase 10.1063/1.121231} {\bibfield  {journal} {\bibinfo
   {journal} {Appl. Phys. Lett.}\ }\textbf {\bibinfo {volume} {72}},\ \bibinfo
  {pages} {1945} (\bibinfo {year} {1998})}\BibitemShut {NoStop}%
\bibitem [{\citenamefont {Sharma}\ \emph {et~al.}(2012)\citenamefont {Sharma},
  \citenamefont {Singh}, \citenamefont {Al-Naib}, \citenamefont {Morandotti},\
  and\ \citenamefont {Ozaki}}]{sharma_terahertz_2012}%
  \BibitemOpen
  \bibfield  {author} {\bibinfo {author} {\bibfnamefont {G.}~\bibnamefont
  {Sharma}}, \bibinfo {author} {\bibfnamefont {K.}~\bibnamefont {Singh}},
  \bibinfo {author} {\bibfnamefont {I.}~\bibnamefont {Al-Naib}}, \bibinfo
  {author} {\bibfnamefont {R.}~\bibnamefont {Morandotti}}, \ and\ \bibinfo
  {author} {\bibfnamefont {T.}~\bibnamefont {Ozaki}},\ }\href {\doibase
  10.1364/OL.37.004338} {\bibfield  {journal} {\bibinfo  {journal} {Opt.
  Lett.}\ }\textbf {\bibinfo {volume} {37}},\ \bibinfo {pages} {4338} (\bibinfo
  {year} {2012})}\BibitemShut {NoStop}%
\bibitem [{\citenamefont {Teo}\ \emph {et~al.}(2015)\citenamefont {Teo},
  \citenamefont {Ofori-Okai}, \citenamefont {Werley},\ and\ \citenamefont
  {Nelson}}]{teo_invited_2015}%
  \BibitemOpen
  \bibfield  {author} {\bibinfo {author} {\bibfnamefont {S.~M.}\ \bibnamefont
  {Teo}}, \bibinfo {author} {\bibfnamefont {B.~K.}\ \bibnamefont {Ofori-Okai}},
  \bibinfo {author} {\bibfnamefont {C.~A.}\ \bibnamefont {Werley}}, \ and\
  \bibinfo {author} {\bibfnamefont {K.~A.}\ \bibnamefont {Nelson}},\ }\href
  {\doibase 10.1063/1.4921389} {\bibfield  {journal} {\bibinfo  {journal} {Rev.
  Sci. Instrum.}\ }\textbf {\bibinfo {volume} {86}},\ \bibinfo {pages} {051301}
  (\bibinfo {year} {2015})}\BibitemShut {NoStop}%
\bibitem [{\citenamefont {Goda}\ \emph {et~al.}(2009)\citenamefont {Goda},
  \citenamefont {Solli}, \citenamefont {Tsia},\ and\ \citenamefont
  {Jalali}}]{goda_theory_2009}%
  \BibitemOpen
  \bibfield  {author} {\bibinfo {author} {\bibfnamefont {K.}~\bibnamefont
  {Goda}}, \bibinfo {author} {\bibfnamefont {D.~R.}\ \bibnamefont {Solli}},
  \bibinfo {author} {\bibfnamefont {K.~K.}\ \bibnamefont {Tsia}}, \ and\
  \bibinfo {author} {\bibfnamefont {B.}~\bibnamefont {Jalali}},\ }\href
  {\doibase 10.1103/PhysRevA.80.043821} {\bibfield  {journal} {\bibinfo
  {journal} {Phys. Rev. A}\ }\textbf {\bibinfo {volume} {80}},\ \bibinfo
  {pages} {043821} (\bibinfo {year} {2009})}\BibitemShut {NoStop}%
\bibitem [{\citenamefont {Goda}\ and\ \citenamefont
  {Jalali}(2013)}]{goda_dispersive_2013}%
  \BibitemOpen
  \bibfield  {author} {\bibinfo {author} {\bibfnamefont {K.}~\bibnamefont
  {Goda}}\ and\ \bibinfo {author} {\bibfnamefont {B.}~\bibnamefont {Jalali}},\
  }\href {\doibase 10.1038/nphoton.2012.359} {\bibfield  {journal} {\bibinfo
  {journal} {Nat. Photon.}\ }\textbf {\bibinfo {volume} {7}},\ \bibinfo {pages}
  {102} (\bibinfo {year} {2013})}\BibitemShut {NoStop}%
\bibitem [{\citenamefont {Roussel}\ \emph {et~al.}(2015)\citenamefont
  {Roussel}, \citenamefont {Evain}, \citenamefont {Le~Parquier}, \citenamefont
  {Szwaj}, \citenamefont {Bielawski}, \citenamefont {Manceron}, \citenamefont
  {Brubach}, \citenamefont {Tordeux}, \citenamefont {Ricaud}, \citenamefont
  {Cassinari}, \citenamefont {Labat}, \citenamefont {Couprie},\ and\
  \citenamefont {Roy}}]{roussel_observing_2015}%
  \BibitemOpen
  \bibfield  {author} {\bibinfo {author} {\bibfnamefont {E.}~\bibnamefont
  {Roussel}}, \bibinfo {author} {\bibfnamefont {C.}~\bibnamefont {Evain}},
  \bibinfo {author} {\bibfnamefont {M.}~\bibnamefont {Le~Parquier}}, \bibinfo
  {author} {\bibfnamefont {C.}~\bibnamefont {Szwaj}}, \bibinfo {author}
  {\bibfnamefont {S.}~\bibnamefont {Bielawski}}, \bibinfo {author}
  {\bibfnamefont {L.}~\bibnamefont {Manceron}}, \bibinfo {author}
  {\bibfnamefont {J.-B.}\ \bibnamefont {Brubach}}, \bibinfo {author}
  {\bibfnamefont {M.-A.}\ \bibnamefont {Tordeux}}, \bibinfo {author}
  {\bibfnamefont {J.-P.}\ \bibnamefont {Ricaud}}, \bibinfo {author}
  {\bibfnamefont {L.}~\bibnamefont {Cassinari}}, \bibinfo {author}
  {\bibfnamefont {M.}~\bibnamefont {Labat}}, \bibinfo {author} {\bibfnamefont
  {M.-E.}\ \bibnamefont {Couprie}}, \ and\ \bibinfo {author} {\bibfnamefont
  {P.}~\bibnamefont {Roy}},\ }\href {\doibase 10.1038/srep10330} {\bibfield
  {journal} {\bibinfo  {journal} {Sci. Rep.}\ }\textbf {\bibinfo {volume}
  {5}},\ \bibinfo {pages} {10330} (\bibinfo {year} {2015})}\BibitemShut
  {NoStop}%
\bibitem [{\citenamefont {Szwaj}\ \emph {et~al.}(2016)\citenamefont {Szwaj},
  \citenamefont {Evain}, \citenamefont {Le~Parquier}, \citenamefont {Roy},
  \citenamefont {Manceron}, \citenamefont {Brubach}, \citenamefont {Tordeux},\
  and\ \citenamefont {Bielawski}}]{szwaj_high_2016}%
  \BibitemOpen
  \bibfield  {author} {\bibinfo {author} {\bibfnamefont {C.}~\bibnamefont
  {Szwaj}}, \bibinfo {author} {\bibfnamefont {C.}~\bibnamefont {Evain}},
  \bibinfo {author} {\bibfnamefont {M.}~\bibnamefont {Le~Parquier}}, \bibinfo
  {author} {\bibfnamefont {P.}~\bibnamefont {Roy}}, \bibinfo {author}
  {\bibfnamefont {L.}~\bibnamefont {Manceron}}, \bibinfo {author}
  {\bibfnamefont {J.-B.}\ \bibnamefont {Brubach}}, \bibinfo {author}
  {\bibfnamefont {M.-A.}\ \bibnamefont {Tordeux}}, \ and\ \bibinfo {author}
  {\bibfnamefont {S.}~\bibnamefont {Bielawski}},\ }\href {\doibase
  10.1063/1.4964702} {\bibfield  {journal} {\bibinfo  {journal} {Rev. Sci.
  Instrum.}\ }\textbf {\bibinfo {volume} {87}},\ \bibinfo {pages} {103111}
  (\bibinfo {year} {2016})}\BibitemShut {NoStop}%
\bibitem [{\citenamefont {Evain}\ \emph {et~al.}(2017)\citenamefont {Evain},
  \citenamefont {Roussel}, \citenamefont {Le~Parquier}, \citenamefont {Szwaj},
  \citenamefont {Tordeux}, \citenamefont {Brubach}, \citenamefont {Manceron},
  \citenamefont {Roy},\ and\ \citenamefont {Bielawski}}]{evain_direct_2017}%
  \BibitemOpen
  \bibfield  {author} {\bibinfo {author} {\bibfnamefont {C.}~\bibnamefont
  {Evain}}, \bibinfo {author} {\bibfnamefont {E.}~\bibnamefont {Roussel}},
  \bibinfo {author} {\bibfnamefont {M.}~\bibnamefont {Le~Parquier}}, \bibinfo
  {author} {\bibfnamefont {C.}~\bibnamefont {Szwaj}}, \bibinfo {author}
  {\bibfnamefont {M.-A.}\ \bibnamefont {Tordeux}}, \bibinfo {author}
  {\bibfnamefont {J.-B.}\ \bibnamefont {Brubach}}, \bibinfo {author}
  {\bibfnamefont {L.}~\bibnamefont {Manceron}}, \bibinfo {author}
  {\bibfnamefont {P.}~\bibnamefont {Roy}}, \ and\ \bibinfo {author}
  {\bibfnamefont {S.}~\bibnamefont {Bielawski}},\ }\href {\doibase
  10.1103/PhysRevLett.118.054801} {\bibfield  {journal} {\bibinfo  {journal}
  {Phys. Rev. Lett.}\ }\textbf {\bibinfo {volume} {118}},\ \bibinfo {pages}
  {054801} (\bibinfo {year} {2017})}\BibitemShut {NoStop}%
\bibitem [{\citenamefont {Steffen}\ \emph {et~al.}(2020)\citenamefont
  {Steffen}, \citenamefont {Gerth}, \citenamefont {Caselle}, \citenamefont
  {Felber}, \citenamefont {Kozak}, \citenamefont {Makowski}, \citenamefont
  {Mavrič}, \citenamefont {Mielczarek}, \citenamefont {Peier}, \citenamefont
  {Przygoda},\ and\ \citenamefont {Rota}}]{steffen_compact_2020}%
  \BibitemOpen
  \bibfield  {author} {\bibinfo {author} {\bibfnamefont {B.}~\bibnamefont
  {Steffen}}, \bibinfo {author} {\bibfnamefont {C.}~\bibnamefont {Gerth}},
  \bibinfo {author} {\bibfnamefont {M.}~\bibnamefont {Caselle}}, \bibinfo
  {author} {\bibfnamefont {M.}~\bibnamefont {Felber}}, \bibinfo {author}
  {\bibfnamefont {T.}~\bibnamefont {Kozak}}, \bibinfo {author} {\bibfnamefont
  {D.~R.}\ \bibnamefont {Makowski}}, \bibinfo {author} {\bibfnamefont
  {U.}~\bibnamefont {Mavrič}}, \bibinfo {author} {\bibfnamefont
  {A.}~\bibnamefont {Mielczarek}}, \bibinfo {author} {\bibfnamefont
  {P.}~\bibnamefont {Peier}}, \bibinfo {author} {\bibfnamefont
  {K.}~\bibnamefont {Przygoda}}, \ and\ \bibinfo {author} {\bibfnamefont
  {L.}~\bibnamefont {Rota}},\ }\href {\doibase 10.1063/1.5142833} {\bibfield
  {journal} {\bibinfo  {journal} {Rev. Sci. Instrum.}\ }\textbf {\bibinfo
  {volume} {91}},\ \bibinfo {pages} {045123} (\bibinfo {year}
  {2020})}\BibitemShut {NoStop}%
\bibitem [{\citenamefont {Roussel}\ \emph {et~al.}(2022)\citenamefont
  {Roussel}, \citenamefont {Szwaj}, \citenamefont {Evain}, \citenamefont
  {Steffen}, \citenamefont {Gerth}, \citenamefont {Jalali},\ and\ \citenamefont
  {Bielawski}}]{roussel_phase_2022}%
  \BibitemOpen
  \bibfield  {author} {\bibinfo {author} {\bibfnamefont {E.}~\bibnamefont
  {Roussel}}, \bibinfo {author} {\bibfnamefont {C.}~\bibnamefont {Szwaj}},
  \bibinfo {author} {\bibfnamefont {C.}~\bibnamefont {Evain}}, \bibinfo
  {author} {\bibfnamefont {B.}~\bibnamefont {Steffen}}, \bibinfo {author}
  {\bibfnamefont {C.}~\bibnamefont {Gerth}}, \bibinfo {author} {\bibfnamefont
  {B.}~\bibnamefont {Jalali}}, \ and\ \bibinfo {author} {\bibfnamefont
  {S.}~\bibnamefont {Bielawski}},\ }\href {\doibase 10.1038/s41377-021-00696-2}
  {\bibfield  {journal} {\bibinfo  {journal} {Light Sci. Appl.}\ }\textbf
  {\bibinfo {volume} {11}},\ \bibinfo {pages} {14} (\bibinfo {year}
  {2022})}\BibitemShut {NoStop}%
\bibitem [{\citenamefont {Couture}\ \emph {et~al.}(2023)\citenamefont
  {Couture}, \citenamefont {Cui}, \citenamefont {Lippl}, \citenamefont {Ostic},
  \citenamefont {Fandio}, \citenamefont {Yalavarthi}, \citenamefont
  {Vishnuradhan}, \citenamefont {Gamouras}, \citenamefont {Joly},\ and\
  \citenamefont {Ménard}}]{couture_single-pulse_2023}%
  \BibitemOpen
  \bibfield  {author} {\bibinfo {author} {\bibfnamefont {N.}~\bibnamefont
  {Couture}}, \bibinfo {author} {\bibfnamefont {W.}~\bibnamefont {Cui}},
  \bibinfo {author} {\bibfnamefont {M.}~\bibnamefont {Lippl}}, \bibinfo
  {author} {\bibfnamefont {R.}~\bibnamefont {Ostic}}, \bibinfo {author}
  {\bibfnamefont {D.~J.~J.}\ \bibnamefont {Fandio}}, \bibinfo {author}
  {\bibfnamefont {E.~K.}\ \bibnamefont {Yalavarthi}}, \bibinfo {author}
  {\bibfnamefont {A.}~\bibnamefont {Vishnuradhan}}, \bibinfo {author}
  {\bibfnamefont {A.}~\bibnamefont {Gamouras}}, \bibinfo {author}
  {\bibfnamefont {N.~Y.}\ \bibnamefont {Joly}}, \ and\ \bibinfo {author}
  {\bibfnamefont {J.-M.}\ \bibnamefont {Ménard}},\ }\href {\doibase
  10.1038/s41467-023-38354-3} {\bibfield  {journal} {\bibinfo  {journal} {Nat.
  Commun.}\ }\textbf {\bibinfo {volume} {14}},\ \bibinfo {pages} {2595}
  (\bibinfo {year} {2023})}\BibitemShut {NoStop}%
\bibitem [{\citenamefont {Gao}\ \emph {et~al.}(2022)\citenamefont {Gao},
  \citenamefont {Zhang}, \citenamefont {Liu},\ and\ \citenamefont
  {Nelson}}]{gao_high-speed_2022}%
  \BibitemOpen
  \bibfield  {author} {\bibinfo {author} {\bibfnamefont {F.~Y.}\ \bibnamefont
  {Gao}}, \bibinfo {author} {\bibfnamefont {Z.}~\bibnamefont {Zhang}}, \bibinfo
  {author} {\bibfnamefont {Z.-J.}\ \bibnamefont {Liu}}, \ and\ \bibinfo
  {author} {\bibfnamefont {K.~A.}\ \bibnamefont {Nelson}},\ }\href {\doibase
  10.1364/OL.462624} {\bibfield  {journal} {\bibinfo  {journal} {Opt. Lett.}\
  }\textbf {\bibinfo {volume} {47}},\ \bibinfo {pages} {3479} (\bibinfo {year}
  {2022})}\BibitemShut {NoStop}%
\bibitem [{\citenamefont {Hebling}\ \emph {et~al.}(2008)\citenamefont
  {Hebling}, \citenamefont {Yeh}, \citenamefont {Hoffmann}, \citenamefont
  {Bartal},\ and\ \citenamefont {Nelson}}]{hebling_generation_2008}%
  \BibitemOpen
  \bibfield  {author} {\bibinfo {author} {\bibfnamefont {J.}~\bibnamefont
  {Hebling}}, \bibinfo {author} {\bibfnamefont {K.-L.}\ \bibnamefont {Yeh}},
  \bibinfo {author} {\bibfnamefont {M.~C.}\ \bibnamefont {Hoffmann}}, \bibinfo
  {author} {\bibfnamefont {B.}~\bibnamefont {Bartal}}, \ and\ \bibinfo {author}
  {\bibfnamefont {K.~A.}\ \bibnamefont {Nelson}},\ }\href {\doibase
  10.1364/JOSAB.25.0000B6} {\bibfield  {journal} {\bibinfo  {journal} {J. Opt.
  Soc. Am. B}\ }\textbf {\bibinfo {volume} {25}},\ \bibinfo {pages} {B6}
  (\bibinfo {year} {2008})}\BibitemShut {NoStop}%
\bibitem [{\citenamefont {Jiang}\ \emph {et~al.}(1999)\citenamefont {Jiang},
  \citenamefont {Sun}, \citenamefont {Chen},\ and\ \citenamefont
  {Zhang}}]{jiang_electro-optic_1999}%
  \BibitemOpen
  \bibfield  {author} {\bibinfo {author} {\bibfnamefont {Z.}~\bibnamefont
  {Jiang}}, \bibinfo {author} {\bibfnamefont {F.~G.}\ \bibnamefont {Sun}},
  \bibinfo {author} {\bibfnamefont {Q.}~\bibnamefont {Chen}}, \ and\ \bibinfo
  {author} {\bibfnamefont {X.-C.}\ \bibnamefont {Zhang}},\ }\href {\doibase
  10.1063/1.123495} {\bibfield  {journal} {\bibinfo  {journal} {Appl. Phys.
  Lett.}\ }\textbf {\bibinfo {volume} {74}},\ \bibinfo {pages} {1191} (\bibinfo
  {year} {1999})}\BibitemShut {NoStop}%
\bibitem [{\citenamefont {Blanchard}\ \emph {et~al.}(2007)\citenamefont
  {Blanchard}, \citenamefont {Razzari}, \citenamefont {Bandulet}, \citenamefont
  {Sharma}, \citenamefont {Morandotti}, \citenamefont {Kieffer}, \citenamefont
  {Ozaki}, \citenamefont {Reid}, \citenamefont {Tiedje}, \citenamefont
  {Haugen},\ and\ \citenamefont {Hegmann}}]{blanchard_generation_2007}%
  \BibitemOpen
  \bibfield  {author} {\bibinfo {author} {\bibfnamefont {F.}~\bibnamefont
  {Blanchard}}, \bibinfo {author} {\bibfnamefont {L.}~\bibnamefont {Razzari}},
  \bibinfo {author} {\bibfnamefont {H.-C.}\ \bibnamefont {Bandulet}}, \bibinfo
  {author} {\bibfnamefont {G.}~\bibnamefont {Sharma}}, \bibinfo {author}
  {\bibfnamefont {R.}~\bibnamefont {Morandotti}}, \bibinfo {author}
  {\bibfnamefont {J.-C.}\ \bibnamefont {Kieffer}}, \bibinfo {author}
  {\bibfnamefont {T.}~\bibnamefont {Ozaki}}, \bibinfo {author} {\bibfnamefont
  {M.}~\bibnamefont {Reid}}, \bibinfo {author} {\bibfnamefont {H.~F.}\
  \bibnamefont {Tiedje}}, \bibinfo {author} {\bibfnamefont {H.~K.}\
  \bibnamefont {Haugen}}, \ and\ \bibinfo {author} {\bibfnamefont {F.~A.}\
  \bibnamefont {Hegmann}},\ }\href {\doibase 10.1364/OE.15.013212} {\bibfield
  {journal} {\bibinfo  {journal} {Opt. Express}\ }\textbf {\bibinfo {volume}
  {15}},\ \bibinfo {pages} {13212} (\bibinfo {year} {2007})}\BibitemShut
  {NoStop}%
\bibitem [{\citenamefont {Naftaly}\ and\ \citenamefont
  {Dudley}(2009)}]{naftaly_methodologies_2009}%
  \BibitemOpen
  \bibfield  {author} {\bibinfo {author} {\bibfnamefont {M.}~\bibnamefont
  {Naftaly}}\ and\ \bibinfo {author} {\bibfnamefont {R.}~\bibnamefont
  {Dudley}},\ }\href {\doibase 10.1364/OL.34.001213} {\bibfield  {journal}
  {\bibinfo  {journal} {Opt. Lett.}\ }\textbf {\bibinfo {volume} {34}},\
  \bibinfo {pages} {1213} (\bibinfo {year} {2009})}\BibitemShut {NoStop}%
\end{thebibliography}%

\end{document}